\documentclass[letterpaper, 10 pt, conference]{ieeeconf}  % Comment this line out
\IEEEoverridecommandlockouts                              % This command is only
\usepackage{graphicx} % for pdf, bitmapped graphics files
\usepackage{times} % assumes new font selection scheme installed
\usepackage{amsmath} % assumes amsmath package installed
\usepackage{amssymb}  % assumes amsmath package installed
\usepackage{color}
\usepackage{cite}
\usepackage{bm}

\title{\LARGE \bf
	Development and Evaluation of Two Learning-Based Personalized Driver Models for Car-Following Behaviors 
}

\author{Wenshuo Wang$^{1}$, {\it Student Member, IEEE}, Ding Zhao$^{2}$, Junqiang Xi$^{3}$, David J. LeBlanc$^{2}$, and J. Karl Hedrick$^{4}$% <-this % stops a space
	\thanks{*This work was supported by China Scholarship Council.}% <-this % stops a space
	\thanks{$^{1}$Wenshuo Wang is with the School of Mechanical Engineering, Beijing Institute of Technology, Beijing, China, 100081, and with  the Department of Mechanical Engineering, University of California Berkeley, CA, 94720 USA.
		{\tt\small wwsbit@gmail.com}}%
	\thanks{$^{2}$Ding Zhao and  David J. LeBlanc are with the University of Michigan Transportation Research Institute, Ann Arbor, MI 48109 
		{\tt\small zhaoding@umich.edu}}%
	\thanks{$^{3}$Junqiang Xi is with the Department of Mechanical Engineering, Beijing Institute of Technology, Beijing, China, 100081.
		{\tt\small xijunqiang@bit.edu.cn}}%
	\thanks{$^{4}$J. Karl Hedrick is with the Department of Mechanical Engineering, University of California at Berkeley, Berkeley, CA 94720 USA.
		{\tt\small karlhed@gmail.com}}%
}

\begin{document}
	
	\maketitle
	\thispagestyle{empty}
	\pagestyle{empty}
	
	%%%%%%%%%%%%%%%%%%%%%%%%%%%%%%%%%%%%%%%%%%%%%%%%%%%%%%%%%%%%%%%%%%%%%%%%%%%%%%%%
	\begin{abstract}
		
		Personalized driver models play a key role in the development of advanced driver assistance systems and automated driving systems. Traditionally, physical-based driver models with fixed structures usually lack the flexibility to describe the uncertainties and high non-linearity of driver behaviors. In this paper, two kinds of learning-based car-following personalized driver models were developed using naturalistic driving data collected from the University of Michigan Safety Pilot Model Deployment program. One model is developed by combining the Gaussian Mixture Model (GMM) and the Hidden Markov Model (HMM), and the other one is developed by combining the Gaussian Mixture Model (GMM) and Probability Density Functions (PDF). Fitting results between the two approaches were analyzed with different model inputs and numbers of GMM components. Statistical analyses show that both models provide good performance of fitting while the GMM--PDF approach shows a higher potential to increase the model accuracy given a higher dimension of training data.
	\end{abstract}
	
	
	%%%%%%%%%%%%%%%%%%%%%%%%%%%%%%%%%%%%%%%%%%%%%%%%%%%%%%%%%%%%%%%%%%%%%%%%%%%%%%%%
	\section{Introduction}
	
	Understanding individual driver behaviors and development of personalized driver models are critical for active safety control systems \cite{lefevre15driver,Butakov15,Butakov16}, vehicle dynamic performance \cite{fu2010modeling}, and human-centered vehicle control systems \cite{wang2015human,wang2014modeling,wang16humancentered}, eco-driving systems \cite{xiang2015closed}, and automated vehicles \cite{Lefevre16}. For instance, a driver assistance system will be more effective if the individual characteristics or/and driving styles can be incorporated \cite{Butakov15}. Personalized driver models can be referred to \cite{Lefevre16} ``\textit{a driver model which can generate the output sequences being as close as possible to what the individual driver would have done in the same driving situation}''.  Lefevre \textit{et al}. \cite{lefevre15driver}, Butakov and Ioannou \cite{Butakov15,Butakov16} developed a personalized driver model based on the  Gaussian mixture model and then applied to the advanced driver assistance systems (ADASs), increasing the potential for more widespread acceptance and use of ADASs.
	
	Generally, the ways to establish a personalized driver model can be grouped into two categories: {\it physical-based model} and {\it learning-based model}. For the physical-based model, formulations with unknown parameters are usually used to describe the structure of driver's driving behaviors such as car following, path following, lane change, overtaking. The major benefit of the physical-based model is that most model parameters have their specific physical meanings, enabling them to be easily interpreted. For example, the intelligent driver model (IDM), optimal velocity model (OVM) \cite{miyajima2007driver}, and control-oriented car-following model are popular physical-based models in the applications of vehicle control \cite{eben2013economy,Zhao2016AcceleratedTechniques,Zhao2016AcceleratedManeuvers} and traffic flow analysis \cite{peng2013new}. The model parameters can be identified using parameter estimation approaches \cite{jin2014reducing,rahman2015improving} such as least squares, Kalman filter, stochastic parameter estimation, etc. The physical-based approach can model driver's basic behavior, however, it is hard to model uncertainties and non-linearity because of the uncertainty and diversity of individual driver's behavior and driving environment. Fortunately, learning-based models can be developed to overcome these issues. Popular approaches have been developed to generate a learning-based driver model such as stochastic switched AutoRegressive eXogenous model (SS-ARX) \cite{sekizawa2007modeling,celik2010predictive}, hidden Markov model \cite{nechyba2001learning}, neural network \cite{Khodayari12,wahab2009driving}, and Gaussian mixture model \cite{wahab2009driving}. These models are believed to represent an individual driver's driving characteristics and describe the underlying source after correctly training. However, it is difficult to explain the physical meaning of the model parameters when learning-based models are directly utilized to generate a highly nonlinear function for driver's behavior (e.g., decision-making and control). Butakov and Ioannou \cite{Butakov15} created a more explainable, flexible, and accurate driver lane change model by combining the learning-based and physical-based methods together, in which physical-based model was used to mimic the driving behavior and the learning-based model (i.e., GMM) was used to describe the parameter distribution of the physical model.  
	
	In the above mentioned learning-based approaches, the GMM method is usually chosen to establish driver model due to its effectiveness of modeling driving tasks \cite{Butakov15,butakov2012driver,angkititrakul2011modeling,miyajima2007driver,Zhao2015AcceleratedData,wang2017learning}. However, limited works studied the comparison between different learning-based approaches. In this paper, two learning-based approaches were shown, in which the influences of different combinations between parameters (e.g., vehicle speed, range, relative speed.) with different numbers of GMM component on model performance were analyzed and discussed. This paper provides the systematic re-examination, evaluation, and comparison of two learning-based approaches for modeling driver's car-following behavior, and also helps researchers understand how many and what parameters are more suitable to model a driver's car-following behavior.
	
	The structure of this paper is organized as follows. Section II shows the problem formulation of driver's car-following behavior. Section III presents the basic methods for personalized driver model. Section IV shows the data collection and data preprocessing. Section V discusses and analyzes the experiment results. 
	
	\section{Problem Formulation}
	\subsection{Personalized Driver Model}
	A specific definition of the personalized driver model is given as: A personalized driver model can be referred to a model that can generate or predict an individual driver's operating parameter (e.g., steering angle, throttle opening, braking force.) or behavior (e.g., lane change, stop \& go, overtaking, decision-making with traffic light.) with the same environment inputs, including traffic users (e.g., other vehicles, bicycles, and pedestrians.), weather conditions, and road conditions.
	
	In this paper, we are going to investigate the personalized driver models for car-following behaviors, which can generate a personalized longitudinal control signal sequence (i.e., acceleration).
	
	\subsection{Car-Following Scenario}
	
\begin{figure}[t]
		\centering
		\includegraphics[width = 0.48\textwidth]{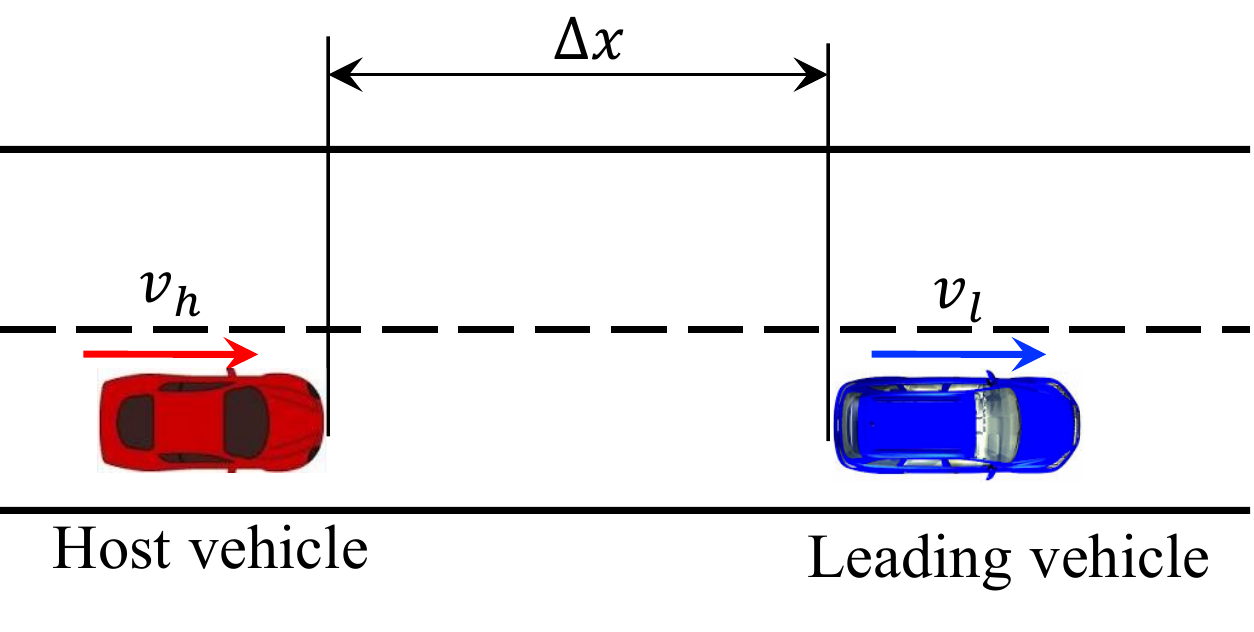}
		\caption{An illustration of the car-following scenario.}
		\label{CarFollowing}
	\end{figure}

	The car-following behavior can be illustrated by Fig. \ref{CarFollowing}. We define the following variables to represent the relative motion of the host vehicle and the vehicle located ahead in the same lane as leading vehicle.
	
	\begin{itemize}
		\item $ \xi_{h} = [x^{h}_{t}, v^{h}_{t} , \jmath_{t}^{h} ]^{\top} \in \mathbb{R}^{3 \times 1}$ is the state of the host vehicle at time $t$, where $x_{t}^{h} \in \mathbb{R}^{+} $ is the longitudinal position of the host vehicle, $v^{h}_{t}$ is the longitudinal speed of the host vehicle, and $ \jmath_{t}^{h}$ is the jerk of the host vehicle  defined as $ \jmath_{t}^{h} =  \ddot{v}^{h}_{t}$.
		\item $ \xi_{l} = [x^{l}_{t}, v^{l}_{t}]^{\top} \in \mathbb{R}^{2 \times 1} $ is the state of the leading vehicle at time $t$, where $x_{t}^{l} \in \mathbb{R}^{+} $ is the longitudinal position of the leading vehicle and $v^{l}_{t}$ is the longitudinal speed of the leading vehicle.
		\item $z_{t} = [\Delta x_{t}, \Delta v_{t}, \Delta \dot{v}_{t}, v^{h}_{t} , \jmath_{t}^{h}]^{\top} \in \mathbb{R}^{5 \times 1}$ are the current states representing current driving situation at time $ t $, where $ \Delta x_{t}  = x^{l}_{t} - x^{h}_{t}$ is the relative distance between the host vehicle and the leading vehicles, $ \Delta v_{t} = v^{l}_{t} - v^{h}_{t}$ is the relative speed between the host vehicle and the leading vehicle, and $ \Delta \dot{v}_{t}$ is the relative acceleration between two vehicles.
	\end{itemize}
	
	The history of explanatory variables, $ z_{1:t} $, and acceleration sequences, $a_{1:t-1}^{h}$, are taken as the model input. The predicted vehicle acceleration is taken as the model output. At each step $ t $, the learned driver model generates an acceleration $ a^{h}_{t} $. The general form of the proposed driver model is presented as
	
	\begin{equation}
	\mathcal{D}(z_{1:t},a_{1:t-1}^{h}) : z_{t} \mapsto \hat{a}^{h}_{t}
	\end{equation}
	The equation (1) is to generate an acceleration with the current input, $z_{t}$, according to the history information, $\boldsymbol{\xi}_{1:t-1} = [z_{1:t-1},a_{1:t-1}^{h}]$, with the prediction step $ \Delta t = 0.1 $ s.
	
	\section{Methods}
	In this section, two learning-based approaches of modeling a personalized driver car-following behavior are discussed, i.e., the Gaussian Mixture Regression with the Hidden Markov Model (GMR-HMM) and the Gaussian Mixture Model with Probability Density Functions (GMM-PDF). To understand the two approaches, the GMM, HMM, and PDF are separately discussed in the following sections.
	
	\subsection{Gaussian Mixture Model}
	A set of $ d $-dimension sequence, $ \boldsymbol{\xi} = \{ \xi_{i} \}_{i = 1}^{N}$ with $ \xi_{i} \in \mathbb{R}^{d\times 1} $, can be encoded in a combination of $ N $ Gaussian models. Assuming that the data in each component of GMM is subject to a Gaussian distribution:
	
	\begin{equation}
	\xi_{i} \sim \mathcal{N}_{i}(\boldsymbol{\mu}_{i}, \bm{\Sigma}_{i})
	\end{equation}
	where $ \boldsymbol{\mu}_i \in \mathbb{R}^{d \times 1}$ and $ \bm{\Sigma}_{i} \in \mathbb{R}^{d \times d}$  is mean and covariance of the $ i $th Gaussian distribution $ \mathcal{N}_{i} $. For all data $ \boldsymbol{\xi} $, it can be encoded by a Gaussian mixture model:
	
	\begin{equation}
	\begin{split}
	\mathcal{P}(\boldsymbol{\xi}; \boldsymbol{\theta}) & = \sum_{i = 1}^{N}\pi_{i} \mathcal{N}_{i}(\boldsymbol{\xi};\boldsymbol{\mu}_{i},\bm{\Sigma}_{i}) \\
	& = \sum_{i = 1}^{N}\pi_{i} \frac{1}{(2\pi)^{d/2} |\bm{\Sigma}_{i}|^{1/2}} \\
	& \ \ \times  \exp \left\lbrace  -\frac{1}{2}(\boldsymbol{\xi} - \boldsymbol{\mu}_{i})^{\top}\bm{\Sigma}^{-1}(\boldsymbol{\xi} - \boldsymbol{\mu}_{i}) \right\rbrace 
	\end{split}
	\end{equation}
	where $ \boldsymbol{\theta} = \{\bm{\mu}_{i}, \bm{\Sigma}_{i}, \pi_{i}\}, i =1,2,\dots,N $; $ \pi_{i} $ is the prior probability and $ \sum_{i=1}^{N} \pi_{i} = 1 $. 
	
	For the car-following model, if we assign $ \boldsymbol{\xi}_{t} = [z_{t}, a_{t}^{h}] $, the joint distribution between $ z_{t} $ and $ a_{t}^{h} $ can be rewritten as
	
	\begin{equation}\label{eq:jointdb}
	\mathcal{P}(z_{t}, a_{t}^{h}; \boldsymbol{\theta}) \sim \sum_{i = 1}^{N} \pi_{i} \mathcal{N}_{i}(z_{t}, a_{t}^{h};\boldsymbol{\mu}_{i}, \bm{\Sigma}_{i})
	\end{equation}
	The parameter $ \boldsymbol{\theta}$ of (\ref{eq:jointdb}) can be estimated by expectation maximization (EM) algorithm \cite{Butakov15}. For the initial value $ (\boldsymbol{\mu}_{0}, \Sigma_{0})$ at iteration step $ s = 0 $, we apply the $ k $-means clustering method to determine $ \boldsymbol{\mu}_{0} $, and then calculate $ \pi_{0} $. Thus, we can obtain the estimated optimal parameter $ \hat{\boldsymbol{\theta}} $ until the log-likelihood function is convergent or meets the maximum iteration steps $ s \geq s_{\max} $, where the optimal objection for the log-likelihood function is formulated as:
	
	\begin{equation}
	\hat{\boldsymbol{\theta}} = \arg \ \underset{\boldsymbol{\theta}}{\max}   \ \mathcal{L}(\boldsymbol{\theta}) = \arg \ \underset{\boldsymbol{\theta}}{\max} \ \log (\mathcal{P(\boldsymbol{\xi} ; \boldsymbol{\theta})})
	\end{equation}
	The number of GMM component can be determined by Bayesian information criterion (BIC). Further, we also discussed the influences of numbers of GMM component on training and tested the model performance.
	
	Our goal is to generate a personalized acceleration sequence based on the learned driver model. With this purpose in mind, two basic approaches are employed and discussed as follows, i.e., HMM and PDF.
	
	\subsection{Hidden Markov Model}
	The joint distribution $ \mathcal{P}(z_{t}, a_{t}^{h}; \boldsymbol{\theta}) $ is encoded to generate the output of the personalized driver model in a continuous HMM of $ N $ states. Here, each component of GMM is treated as a state of HMM. The HMM can be presented by $\mathcal{H}(\bm{\Pi}, \bm{\Phi}, \bm{\mu}, \bm{\Sigma})$, where $ \bm{\Pi} = \{ \pi^{s=0}_{i} \}_{i=1}^{N} $ is the initial prior probability of being in state $ i $, $ \Phi = \{\phi_{i,j}\}_{i,j}^{N} $ is the transitional probability from state $ i $ to $ j $; $ \mu_{i} $ and $ \bm{\Sigma}_{i} $ are the mean and the covariance matrix of the $ i $th Gaussian distribution of the HMM. Therefore, the input and output components in each state of the HMM are defined as:
	
	\begin{equation}
	\bm{\mu}_{i} = [\mu^{z}_{i}, \mu^{a^{h}}_{i}]^{\top},
	\end{equation}
	\begin{equation}
	\bm{\Sigma}_{i} = 
	\begin{bmatrix}
	\bm{\Sigma}^{z}_{i} & \bm{\Sigma}^{za^{h}_{t}}_{i} \\
	\bm{\Sigma}^{a^{h}z}_{i} & \bm{\Sigma}^{a^{h}}_{i}
	\end{bmatrix},
	\end{equation}
	As such, the acceleration at time $ t $ can be estimated, given the history information, $ \xi_{1:t-1} = [z_{1:t-1}, a^{h}_{1:t-1}]$ and the observed state $ z_{t} $ at time $ t $, by using
	
	\begin{equation}
	\hat{a}^{h}_{t} = \sum_{i = 1}^{N} \alpha_{i}(z_{t})\left[  \mu^{a^{h}_{t}}_{i} + \bm{\Sigma}^{a^{h}_{t}}_{i} (\bm{\Sigma}^{z_{t}}_{i})^{-1} (z_{t} - \mu^{z_{t}}_{i}) \right] 
	\end{equation}
	where $ \alpha_{i}(z_{t}) $ is the HMM forward variable, computed as the probability of being in state $ i $ at time $ t $, given by:
	
	\begin{equation}
	\alpha_{i}(z_{t}) = \frac{\left(  \sum_{j = 1}^{N} \alpha_{j}(z_{t-1})\cdot \phi_{j,i} \right) \cdot \mathcal{N}_{i}(z_{t};\mu^{z}_{i},\bm{\Sigma}^{z}_{i})}{\sum_{l = 1}^{N}\left(  \sum_{j = 1}^{N} \alpha_{j}(z_{t-1})\cdot \phi_{j,i} \right) \cdot \mathcal{N}_{l}(z_{t};\mu^{z}_{l},\bm{\Sigma}a^{z}_{l})}
	\end{equation}
	Here, the initial value at time $ t = 1 $ is computed by 
	\begin{equation*}
	\alpha_{i}(z_{1}) = \frac{\pi_{i}\mathcal{N} (z_{1}; \mu^{z}_{i}, \bm{\Sigma}^{z}_{i})}{\sum_{k = 1}^{N}\pi_{k}\mathcal{N} (z_{1}; \mu^{z}_{k}, \bm{\Sigma}^{z}_{k})}
	\end{equation*}
	
	\subsection{Probability Density Function}
	The second approach to get the estimated output, $ \hat{a}^{h}_{t} $ is to compute the value that can maximize the probability based on the probability density function of the GMM, i.e.,
	
	\begin{equation}
	\hat{a}^{h}_{t} = \arg \ \underset{a^{h}\in \mathcal{A}^{h}}{\max} \ \mathcal{P}(z_{t}, a^{h}; \hat{\boldsymbol{\theta}})
	\end{equation}
	where $ \mathcal{A}^{h} $ is the set of possible value that $ a^{h} $ can reach and $ \hat{\boldsymbol{\theta}} $ is the estimated parameter of the GMM using the collected driving data on the basis of (5).
	
	\section{Experiments for Data Collection}
	In this section, the data collection and the procedure of data training and test are presented.
	\subsection{Data Collection}
	The data used in this paper is from the Safety Pilot Model Deployment (SPMD) database \cite{Bezzina2014}.  It recorded naturalistic driving of 2,842 equipped vehicles in Ann Arbor, Michigan for more than two years. In the SPMD program, 98 sedans are equipped with data acquisition system and MobilEye$^\circledR $ \cite{Zhao2016AcceleratedManeuvers,harding2014vehicle}, which provides: a) relative position to the lead vehicle (range), and b) lane tracking measures about the lane delineation both from the painted boundary lines and the road edge. The error of range measurement is around 10\% at 90 m and 5\% at 45 m \cite{Stein2003}.
	Data in two separate months, October 2012 and April 2013, were downloaded from the U.S. Department of Transportation website \cite{RDEDataEnvironment}. To ensure consistency of the used dataset, we apply the following criteria to extracting the car-following events from the entire datasets:
	
	\begin{itemize}
		\item $\Delta x$ $\in$ [0.1 m, 120 m]
		\item  Longitude $\in$ [−88.2, −82.0]
		\item  Latitude $\in$ [41.0, 44.5]
		\item Duration of car-following $>$ 50 s
	\end{itemize}
	All the car-following events were detected from 76 drivers. To the end, the number of entire purified car-following events is 5,294.
	
	\subsection{Data Training Process}
	\subsubsection{Preprocessing}
	For the $ j $th driver, all the raw data, $ \boldsymbol{\xi}^{j} $, were smoothed by a moving average filter with a window size $ W = 10 $. The data for each single driver were evenly divided into $ M $ groups and then $ M-1 $ groups were randomly selected as the training data and the remaining group was used to test the model, which is also called the {\it leave-one-out cross-validation} method. Here, all the divided data groups for each single driver meet the following conditions:
	
	\begin{equation}
	\underset{p=1}{\bigcup} \boldsymbol{\xi}^{j,p}  = \boldsymbol{\xi}^{j} \ \mathrm{and} \ \underset{p=1}{\bigcap} \boldsymbol{\xi}^{j,p}  = \emptyset, \ \mathrm{with} \ p = 1,2, \cdots, M,
	\end{equation}
	where $ \boldsymbol{\xi}^{j,p} $ presents the $ p^{\mathrm{th}} $ group of data for the $ j^{\mathrm{th}} $ driver, $ \bigcup $ and $ \bigcap $ are union and intersection, respectively; $ \emptyset $ is the empty set. In this paper, we set $ M = 20 $.
	
	\subsubsection{Dimension of Model Inputs}
	We will investigate the influence of different inputs on the model performance. For the personalized driver model, the different input variables are tested using the following combinations:
	
	\begin{itemize}
		\item $ z^{(1)}_{t} = [\Delta x_{t}, \Delta v_{t}]$; $ \boldsymbol{\xi}^{(1)} = [z_{t}, a^{h}_{t}]^{\top} \in \mathbb{R}^{3 \times 1} $;
		\item $ z^{(2)}_{t} = [\Delta x_{t}, \Delta v_{t}, v_{t}^{h}]$; $ \boldsymbol{\xi}^{(2)} = [z_{t}, a^{h}_{t}]^{\top} \in \mathbb{R}^{4 \times 1}$;
		\item $ z^{(3)}_{t} = [\Delta x_{t}, \Delta v_{t}, \Delta \dot{v}_{t}, v_{t}^{h}]  $; $ \boldsymbol{\xi}^{(3)} = [z_{t}, a^{h}_{t}]^{\top} \in \mathbb{R}^{5 \times 1}$;
		\item $ z^{(4)}_{t} = [\Delta x_{t}, \Delta v_{t}, \Delta \dot{v}_{t}, \jmath_{t}^{h}, v_{t}^{h}]$; $ \boldsymbol{\xi}^{(4)} = [z_{t}, a^{h}_{t}]^{\top} \in \mathbb{R}^{6 \times 1} $.
	\end{itemize}
	where $ z^{(i)}_{t}, i = 1,2,3,4, $ represents the $i$th input.  Here, we default that the host vehicle speed, $ v^{h}_{t} $, and relative range, $ \Delta x_{t} $, at current time $ t $ are the basic parameters for describing a driver's car-following behavior. In the training procedure, the training data are $ \boldsymbol{\xi}_{1:t-1} = [z_{1:t-1}, a^{h}_{1:t-1}]$.
	
	\subsubsection{Number of the GMM Components}
	Different numbers of the GMM components will affect the model accuracy. More components will cause the over-fitting problem, and fewer components could not characterize the underlying sources of data and will increase the prediction error. Therefore,  $ N \in $  \{2, 3, 4, 5, 6, 7, 8, 9, 10, 11, 12, 15, 20, 25\} are selected to investigate the influences of the GMM components on model performance.
	
	\subsection{Data Testing Process}
	\begin{figure}[t]
		\centering
		\includegraphics[width  = 0.48\textwidth]{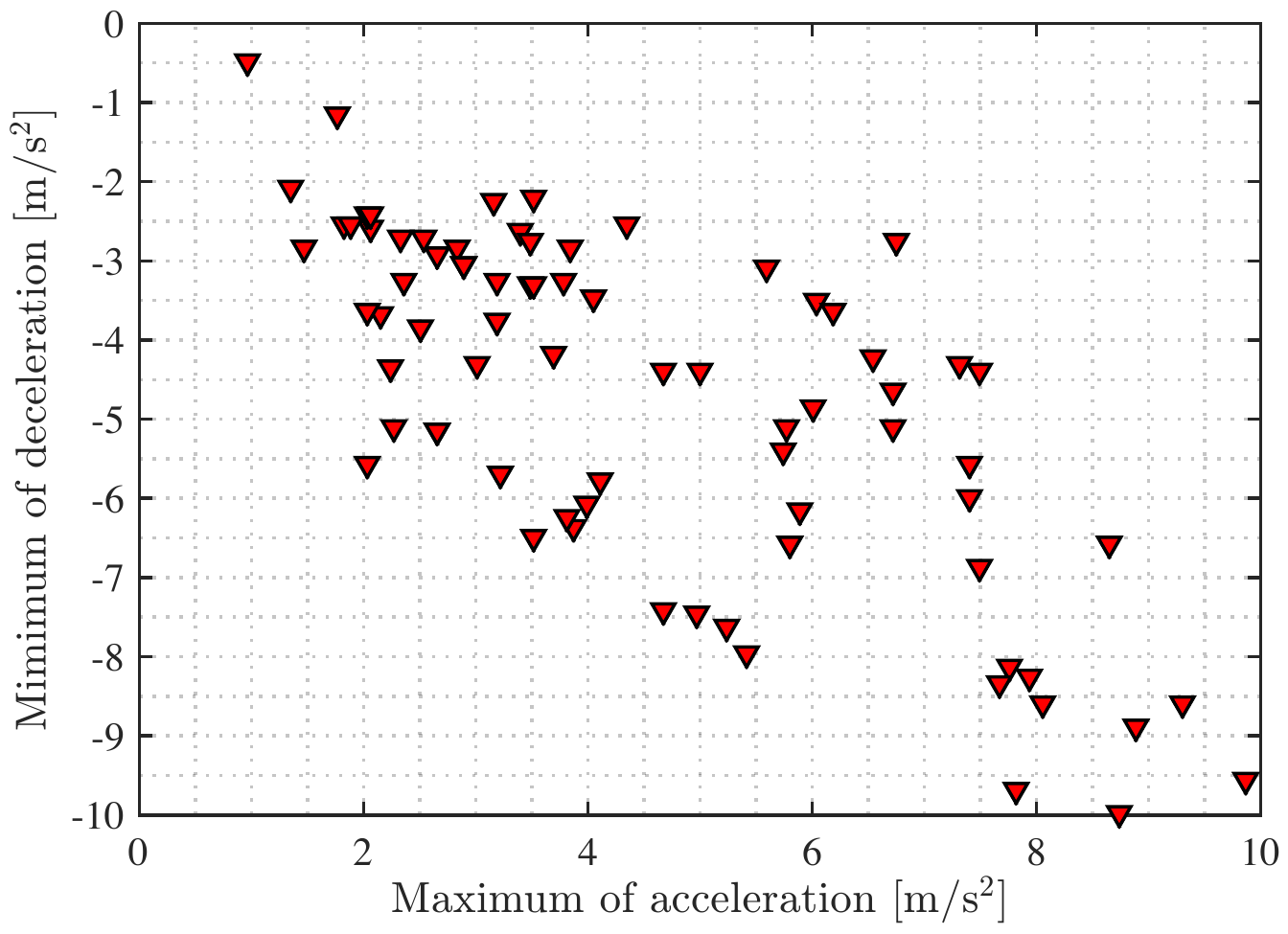}
		\caption{The example of maximum and minimum accelerations for 75 drivers in our experiments.}
		\label{max_min_ax}
	\end{figure}
	
	We will repeatedly run 10 times for each training dataset of a driver participant and the average errors of 10 runs is selected as the performance index to evaluate the model performance. We run 10 times for each training dataset is because the initial value used in (9) is generated by using $ k $-means cluster ($ k $-MC) method in which the initial value was randomly generated.
	
	For the reachable region, $ \mathcal{A}^{h} $ in (10),  we set $ \mathcal{A}^{h} = \{ a^{h} | a^{h}_{\min} \leq a^{h} \leq a^{h}_{\max} \} $. The $ a^{h}_{\min} $ and $ a^{h}_{\max} $ can be generated from the statistical information of each driver, as shown in  Fig. \ref{max_min_ax}. In Fig. \ref{max_min_ax}, 75 driver participants are included and each point represents a driver. For most drivers, the $ a^{h}_{\min} $ and $ a^{h}_{\max} $ are located at  $ [-8,8] $ m/s$ ^{2} $. Therefore, in our work, for all drivers we set $ a^{h}_{\min} = -8  $ m/s$ ^{2} $ and $ a^{h}_{\max} = +8 $ m/s$ ^{2} $. Therefore, when inputing $ z_{t} $, we can obtain an locally optimal corresponding estimated output $ \hat{a}^{h}_{t} $ using (10). 
	
	\subsection{Performance Index}
	The average errors, $ \bar{e} $, between the real value ($ a^{h}_{t} $) and the estimated value ($ \hat{a}^{h}_{t} $) are used as the performance index to evaluate the presented methods and computed by
	
	\begin{equation}
	\begin{split}
	\bar{e}  & = \frac{1}{t_{end}}\int_{0}^{t_{end}} e(\tau) d\tau \\
	& =  \frac{1}{t_{end}}\int_{0}^{t} |\hat{a}^{h}_{\tau} - a^{h}_{\tau}| d\tau
	\end{split}
	\end{equation}
	where $ t_{end} $ is the length of time-indexed test data. A smaller (larger) value of $ \bar{e} $ indicates a better (undesirable) performance for the proposed approaches.
	
	\section{Results Analysis}
	In this section, the training and test results with respect to different input variables and numbers of GMM component based on two approaches, i.e., GMM+HMM and GMM+PDF,  are presented and discussed. To simplify the description and show more clear, we take one of 75 driver participants for example.
	
	\begin{figure}[t]
		\centering
		\includegraphics[width = 0.48\textwidth]{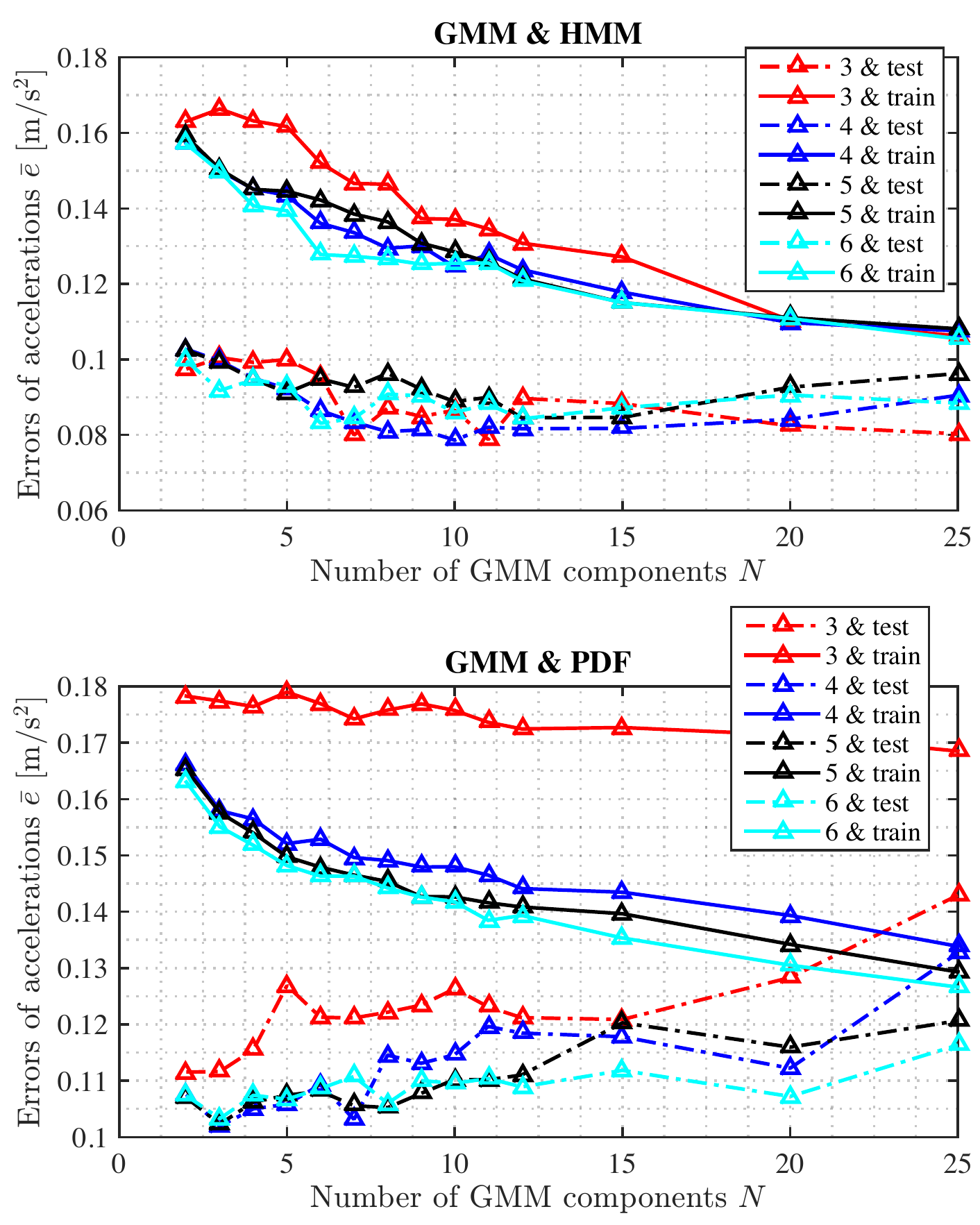}
		\caption{The training errors and test errors for GMM+HMM approach and GMM+PDF approach with different input dimensions.}
		\label{Comp0}
	\end{figure}
	\subsection{Influence of the GMM Component}
	For the different number of GMM components, the training and test accuracy of the model will be different. More components will decrease the training errors, but can result in over-fitting problems and increase computational costs; inversely, fewer components can reduce computational efforts but may induce larger errors. For example, Fig. \ref{Comp0} shows the average errors of training and test results with different numbers of GMM components using different approaches for a driver. The horizontal and vertical axis are the number of GMM component and average errors of acceleration, respectively. The number represents the dimension of training data, as discussed in Section IV, {\it B}. For example, ``5 \& train'' represents the dimension of training input is 5, i.e.,  $ \boldsymbol{\xi}  = [\Delta x_{t}, \Delta v_{t}, \Delta \dot{v}_{t}, v_{t}^{h}, a^{h}_{t}] = [z_{t}, a^{h}_{t}]^{\top} \in \mathbb{R}^{5 \times 1} $, and, correspondingly, ``5 \& test'' represents the input dimension of test data is 4, i.e., $ z_{t} = [\Delta x_{t}, \Delta v_{t}, \Delta \dot{v}_{t}, v_{t}^{h}] $.
	
	\subsubsection{GMM+HMM}
	Top plot in Fig. \ref{Comp0} shows the training and test average errors of acceleration using the GMM+HMM approach.  It is obviously that the training errors are decreasing with the  number of GMM components increasing. The test errors are decreasing with the number of GMM components increasing from 2 to 10; after that, the test errors are increasing slightly.
	
	\subsubsection{GMM+PDF}
	Similarly, the bottom plot in Fig. \ref{Comp0} shows the training and test errors of acceleration using GMM-PDF approach. It can be concluded that the training errors decreases and the test errors of acceleration increases while the number of GMM increases.
	
	\subsection{Influence of Model Inputs}
	
	\subsubsection{GMM+HMM}
	From the top plot of Fig. \ref{Comp0}, we can know that for different kinds of input by using GMM+HMM approach, the training errors are decreasing with a higher dimension of input, but not for the test errors. In addition, for the test results using GMM+HMM approach while the dimension of training data is 4, i.e., $ \boldsymbol{\xi} =  [\Delta x_{t}, \Delta v_{t}, v_{t}^{h}, a^{h}_{t}]^{\top} $, we found that the estimation accuracy is better than others.
	
	\subsubsection{GMM+PDF}
	From the bottom plot of Fig. \ref{Comp0}, it can be seen that for different dimensions of training data with the GMM+PDF approach, the training errors are decreasing with the dimension of training data increasing, and the same case occurs for the test errors. For the GMM+PDF approach, the estimation accuracy is the best when the 6-dimension of training data is chosen, i.e., $ \boldsymbol{\xi}  = [\Delta x_{t}, \Delta v_{t}, \Delta \dot{v}_{t}, \jmath_{t}^{h}, v_{t}^{h}, a^{h}_{t}]^{\top} $.
	
			\begin{figure}[t]
				\centering
				\includegraphics[width = 0.48\textwidth]{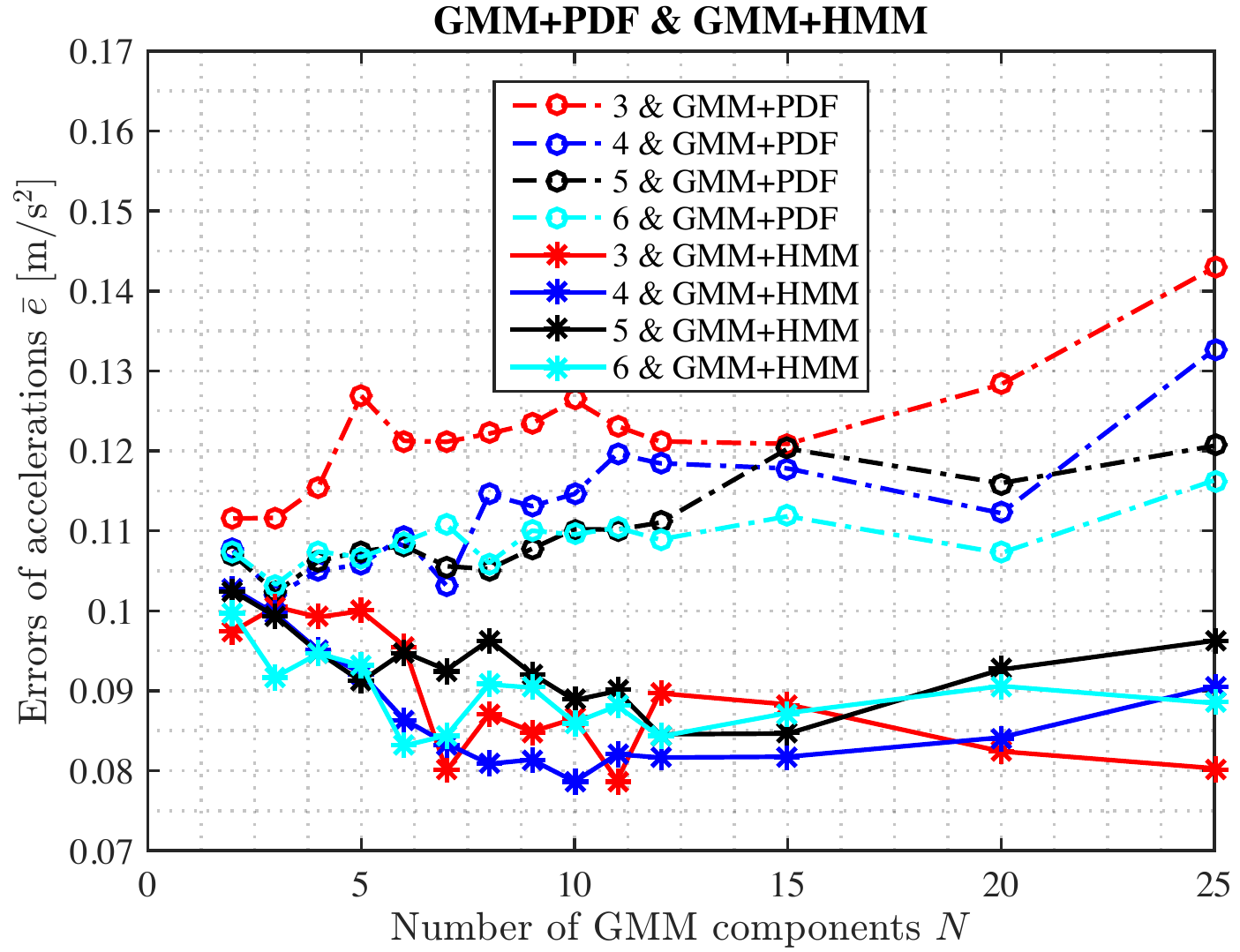}
				\caption{The comparison between the GMM+HMM and GMM+PDF approaches with different input dimensions.}
				\label{comparison1}
			\end{figure}
			
			\begin{figure}[t]
				\centering
				\includegraphics[width = 0.48\textwidth]{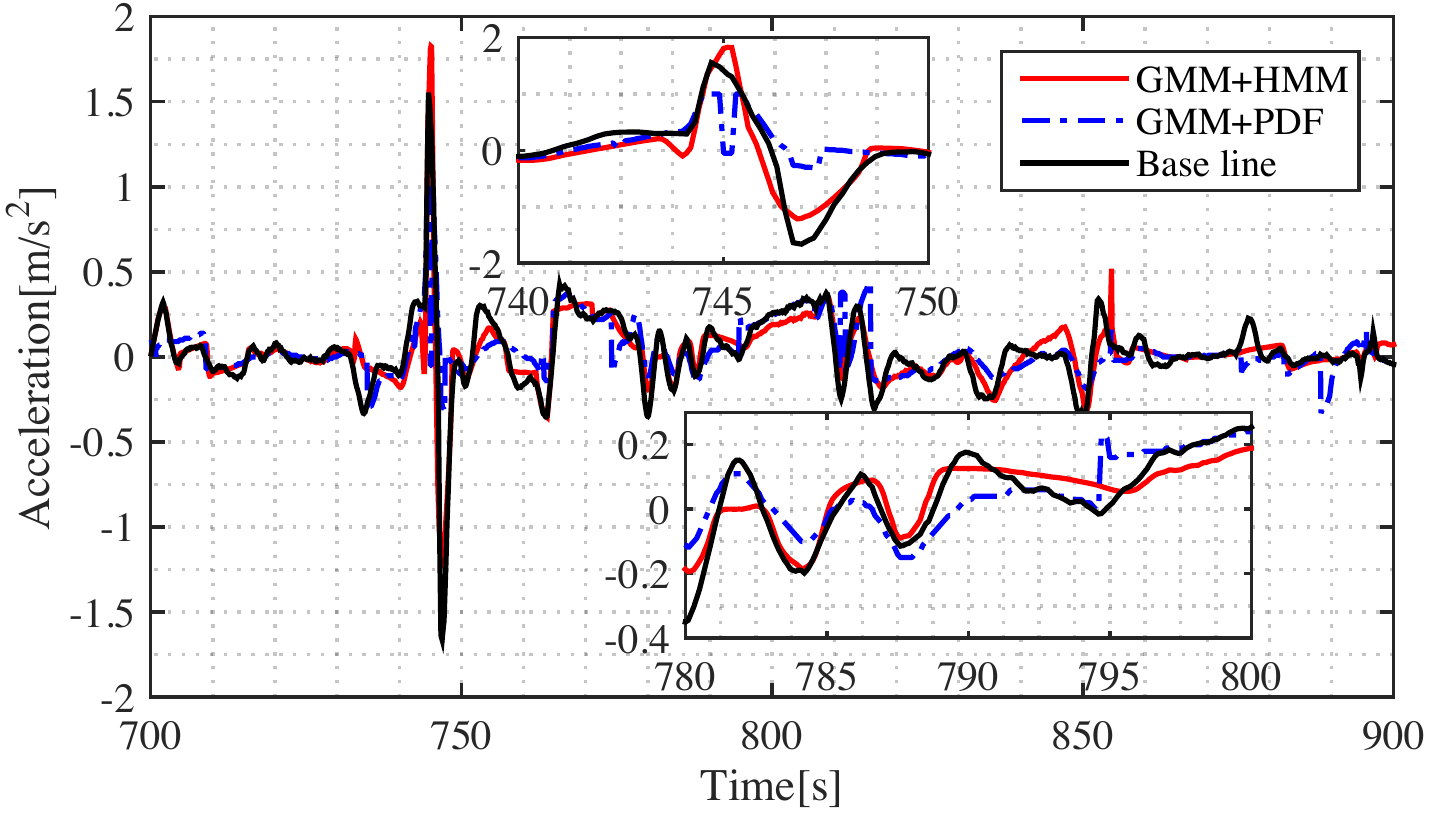}
				\caption{The comparison of acceleration prediction between the GMM+HMM and GMM+PDF approaches with $ N = 12$ GMM components and 4 input variables.}
				\label{comparison2}
			\end{figure} 
	
	\subsection{Comparison Between Two Methods}
	The comparison results between two methods are shown in Fig. \ref{comparison1}. It is obvious that for different dimensions of training data (i.e., $ \boldsymbol{\xi} \in \mathbb{R}^{d \times 1}$, $ d = 3,4,5,6 $), the GMM+HMM approach has a higher estimation accuracy than the GMM+PDF approach. For the GMM+HMM method, the mean estimation errors, $ \bar{e} $, can be lower than 0.1, but for the GMM+PDF method, $ \bar{e} $ is always larger than 0.1, even for different numbers of the GMM components and  dimensions of training data.
	
	Fig. \ref{comparison2} shows the estimation results with two different methods when the dimension of training data is 4 and the number of components is 12. We note that the GMM + PDF method has a higher potential to increase the model accuracy given a higher dimension of training data. 
	
	\section{Conclusions and Future Works}
	
	This paper proposed and compared two personalized driver models in car-following scenarios. The GMM+HMM method (Gaussian mixture model + hidden Markov model) and the GMM+PDF method (Gaussian mixture model + probability density function) were used to fit large-scale naturalistic driving data to describe the uncertainties and nonlinearities of the human behaviors. Different GMM components and training data dimensions was tested out and their influences on the model accuracy were analyzed. For training a personalized car-following driver model, we found that:
	
	\begin{itemize}
		\item For the GMM + HMM method, a higher dimension of the training data might not result in a higher estimation accuracy. The preferred number of the GMM components is 10 $ \sim $ 15 and the preferred dimension of training data is 4, including host vehicle speed, relative range, relative speed, and the acceleration of the host vehicle.
		\item For GMM + PDF methods, a higher dimension of the training data can slightly reduce the estimation errors of acceleration but will increase the computational cost. 
		\item In the car-following case, the GMM + HMM method can catch the underlying sources of naturalistic driving data and shows a better prediction performance than GMM + PDF method by about 27.3\%. 
	\end{itemize}
	
	The Gaussian mixture model is a popular tool to generate a statistical model due to its flexibility and simplicity for learning, but it is sensitive to outliers especially with small numbers of data points. Also, due to the bounded nature of driving behaviors, tails of the Gaussian distributions might be shorter than required, which affects the fitting accuracy. In the future work, we will take the bounded feature of driver behaviors into consideration and develop a learning-based bounded driver model.
	
	
	%%%%%%%%%%%%%%%%%%%%%%%%%%%%%%%%%%%%%%%%%%%%%%%%%%%%%%%%%%%%%%%%%%%%%%%%%%%%%%%%

	%%%%%%%%%%%%%%%%%%%%%%%%%%%%%%%%%%%%%%%%%%%%%%%%%%%%%%%%%%%%%%%%%%%%%%%%%%%%%%%%
	
	%%%%%%%%%%%%%%%%%%%%%%%%%%%%%%%%%%%%%%%%%%%%%%%%%%%%%%%%%%%%%%%%%%%%%%%%%%%%%%%%
	%\section*{APPENDIX}
	
	%\section*{ACKNOWLEDGMENT}

	%%%%%%%%%%%%%%%%%%%%%%%%%%%%%%%%%%%%%%%%%%%%%%%%%%%%%%%%%%%%%%%%%%%%%%%%%%%%%%%%

	\bibliographystyle{IEEEtran}
	
	\bibliography{Bib_Conf}

\end{document}